\documentclass[aps,prd,twocolumn,groupedaddress,showpacs]{revtex4}
\usepackage{graphicx}
\usepackage{dcolumn}
\usepackage{bm}

\begin{document}

\title{Cosmological constant, semiclassical gravity, 
and foundations of quantum mechanics}

\author{Hrvoje Nikoli\'c}
\affiliation{Theoretical Physics Division, Rudjer Bo\v{s}kovi\'{c}
Institute,
P.O.B. 180, HR-10002 Zagreb, Croatia.}
\email{hrvoje@thphys.irb.hr}

\date{\today}

\begin{abstract}
The old cosmological-constant (CC) problem indicates an inconsistency
of the usual formulation of semiclassical gravity.
The usual formulation of semiclassical gravity 
also seems to be inconsistent with the conventional
interpretation of quantum mechanics based on the discontinuous
wave-function collapse.
By reformulating semiclassical gravity in terms of Bohmian
deterministic particle trajectories, the resulting semiclassical 
theory avoids both the old CC problem and the discontinuous
collapse problem of the usual semiclassical theory.
The relevance to the new CC problem and to particle creation 
by classical gravitational fields is also discussed.
\end{abstract}


\maketitle

\section{Introduction}

\subsection{Problems with semiclassical gravity}

As the correct theory of quantum gravity is not yet known,
there is some hope that at least a semiclassical approximation
could work. In this approximation, gravity is treated classically, 
while all other forms of matter are quantized. The semiclassical
theory is usually formulated as a semiclassical Einstein equation
\begin{equation}\label{e1}
G_{\mu\nu}(x)=8\pi G_{\rm N}\langle\Psi|\hat{T}_{\mu\nu}(x)|\Psi\rangle ,
\end{equation}
where $G_{\mu\nu}$ is the Einstein tensor, $G_{\rm N}$ is the 
Newton constant, $\hat{T}_{\mu\nu}$ is a quantum operator
representing the symmetric energy-momentum tensor of matter, 
and $|\Psi\rangle$ is the quantum state.
However, as $\hat{T}_{\mu\nu}$ is calculated from 
quantum field theory (QFT), it contains a huge contribution
from the vacuum energy of the field, leading to a huge contribution
to the cosmological constant, many orders of 
magnitude larger than the measured one. This represents
the core of the cosmological-constant (CC) problem. In the old 
formulation of the problem \cite{weinberg,nobb} one would like 
to find a theoretical mechanism that
makes this vacuum contribution to the cosmological constant 
vanishing, while in the new, more ambitious, formulation of the 
problem \cite{sahni,carroll,padm} one would like to explain 
why the sum of all possible contributions to the cosmological
constant, including that of the vacuum energy, is of the same 
order of magnitude as the matter density of the universe.  

Another, seemingly unrelated, problem with the semiclassical
equation (\ref{e1}) concerns the fundamental interpretational
problems of quantum mechanics (QM) itself. When $|\Psi\rangle$ 
in (\ref{e1}) is a superposition of two macroscopically 
distinct states, then experiments show that (\ref{e1})
is wrong \cite{page}; the measured gravitational field is not 
given by the average value of the energy-momentum in the 
superposition $|\Psi\rangle$, but rather by the actual
measured value of the energy-momentum. One could take this effect
into account by reformulating (\ref{e1}) in terms of a 
quantum state $|\Psi(t)\rangle$, in which the extra time dependence 
corresponds to quantum ``collapses" of $|\Psi\rangle$ induced
by quantum measurements. However, according to the standard
interpretation of QM, the ``collapses" are discontinuous
processes that change $|\Psi\rangle$ instantaneously and nonlocally. 
Consequently, owing to the extra time dependence, 
the energy-momentum in (\ref{e1}) 
ceases to be a smooth function, which implies that it cannot 
satisfy the local conservation equation 
$\nabla^{\mu} \langle\Psi(t)|\hat{T}_{\mu\nu}(x)|\Psi(t)\rangle =0$.
On the other hand, the left-hand side is a classical quantity
that satisfies $\nabla^{\mu}G_{\mu\nu}(x)=0$, suggesting 
an inconsistency of (\ref{e1}).
We refer to this problem as the 
{\em discontinuous collapse problem}.
   
Both problems with the semiclassical equation (\ref{e1})
indicate that the semiclassical approximation is not 
an appropriate framework to deal with interactions 
between gravity and matter. For that reason, it is
very likely that, in order to have a consistent theory,
gravity must also be quantized. Nevertheless, 
we believe that it is too early to completely 
give up the attempts to construct a 
satisfying semiclassical theory that avoids the 
problems outlined above. The aim of this work 
is just to propose such a reformulated semiclassical theory
that avoids these problems.     

\subsection{Main ideas for a solution}

To avoid the discontinuous collapse problem,
we first need to replace the usual notion of instantaneous
discontinuous
wave-function collapse in QM with something smooth and 
continuous. Fortunately, there already exists such a 
formulation of QM - the {\it Bohmian formulation} 
\cite{bohm,bohmPR1,bohmPR2,holrep,holbook}.
(For a comparison with other formulations, see also \cite{ajp}.)
In the case of a completely quantum description of a physical system,
the Bohmian formulation of QM, just as any other formulation,
leads to the {\em same statistical predictions}
as the usual formulation. Nevertheless, in general, a theoretical concept
of a ``semiclassical approximation" is somewhat ambiguous, so 
different approaches to a semiclassical approximation may not be
equivalent. In fact, among various formulations of QM \cite{ajp},
the Bohmian formulation is the most similar to classical mechanics,
so it seems reasonable that the Bohmian approach could be 
the most suitable for a satisfying formulation of a
semiclassical approximation. Besides, the Bohmian interpretation
of quantum gravity \cite{gol,pin1,pin2,shoj} 
has already been found useful
for certain cosmological applications \cite{bar,col,pn,mar,pn2}. Therefore, 
we base our semiclassical formulation of gravity interacting
with matter on a Bohmian description of quantum matter.

In the case of first quantization of particles, the Bohmian 
interpretation assumes that particles are pointlike 
objects with continuous and deterministic trajectories.
However, the force on the particle depends on the 
wave function, which makes these trajectories different
from the classical ones. The particle positions at 
any time are completely determined by the initial conditions.
However, if an observer is ignorant of the actual
initial particle positions, one completely restores
the effective standard probabilistic rules of QM.
Although this hidden-variable formulation of QM is conceptually appealing
and consistent with observations,
most physicists do not use the Bohmian formulation in practice,
mainly because it is technically more complicated than the
standard formulation, with the same measurable statistical predictions
for purely quantum systems.
However, the application of the 
Bohmian formulation to a {\em semiclassical} approximation
may lead to measurable predictions that cannot be 
obtained with other formulations.     
 
In the case of QFT, the Bohmian formulation is constructed
in an analogous way, but with the crucial difference 
that now the fundamental objects having a continuous and deterministic
dependence on time are not pointlike particles, but continuous 
fields. Indeed, in high-energy physics, the dominating point of
view is that the fundamental quantized objects are not particles but fields.
Still, many phenomenologically oriented
particle physicists view QFT merely as a mathematical 
tool useful only for calculation of properties of particles.
Moreover, it seems that it is possible to construct a consistent
particle-scattering formalism that completely avoids any 
referring to fields \cite{schub}.
In fact, there is no real proof that fields (or particles)
are more fundamental objects than particles (or fields) \cite{nikolmyth}.  
In the Bohmian formulation, where particles or fields are 
supposed to objectively exist even when they are not measured,
the field-or-particle dilemma is even sharper than that in the 
standard formulation. To reproduce all good results of both
nonrelativistic first quantization and relativistic QFT, 
in the Bohmian formulation it can be assumed 
that both particles and fields exist separately, such that,
in particle-physics experiments,
particles are objects that are really observed, 
whereas fields play a role in governing continuous deterministic
processes of particle creation and destruction 
\cite{nikfpl1,nikfpl2}.    

If both particles and fields exist separately, then,
in the Bohmian formulation, both particles and fields
generate separate continuously and deterministically 
evolving energy-momentum tensors. 
However, the total energy-momentum tensor cannot be 
a sum of these two tensors, because it would correspond
to a double counting. Instead, either only particles or 
only fields determine the energy-momentum tensor 
on the right-hand side of a
semiclassical Einstein equation. Is that the energy-momentum
of fields, or that of particles? While it is difficult 
to answer this question by using purely theoretical
arguments, it is important to notice the following
essential difference between these two choices:
{\it Whereas the field energy-momentum contains an 
infinite (or huge) vacuum contribution, the particle
energy-momentum does not contain this vacuum contribution 
at all.} Of course, particles in an external potential
may also have a nonzero ground-state energy, but 
such a particle ground-state energy is finite and usually small.
The huge vacuum energy-momentum can be removed 
for fields as well, e.g., by normal ordering, but such a removal
is theoretically artificial. On the other hand, by 
assuming that fundamental objects that determine 
the energy-momentum tensor are not fields but particles,
the vacuum contribution removes automatically. 
This is how the quantum theory formulated in terms of 
Bohmian particle trajectories 
avoids two fundamental problems of (\ref{e1})
at the same time:
the discontinuous collapse problem and the old CC problem.     
 
Before presenting details of such a Bohmian 
formulation, the following remarks are in order.
First, it is often claimed that the existence of the 
Casimir effect is a proof that the vacuum energy is real, 
so that it is unphysical to ignore it. However, 
the fact is that the Casimir effect can be 
derived even without referring to vacuum energy \cite{jaffe}, 
so the existence of the Casimir effect cannot really be 
taken as a proof that vacuum energy is physical.
Instead, the Casimir force can be treated as a 
van der Waals-like force, the enegy-momentum of which
can be described by a 
small potential between {\em real particles} that constitute
the conductive plates.

Second, in curved spacetime, which the semiclassical 
theory of gravity is supposed to describe, 
QFT particle states cannot be defined in a unique way 
\cite{bd}, which is a problem for a theory with an 
ambition to deal with particles as fundamental objectively
existing entities. However, this problem can be avoided
by an introduction of a preferred frame that allows
to define particles in an objective and local-covariant 
manner \cite{nikolcur}. Moreover, it is possible that
a preferred frame is generated dynamically in a covariant way 
(for a concrete proposal see \cite{nikddw}), 
which, at least, makes the idea of a preferred frame less
unpleasant.  

In the next section we formulate the theory with first 
quantization of particles, while the effects of QFT,
including the effects of particle creation and destruction,
are studied in Sec.~\ref{QFT}. Some further physical implications,
including the relevance to the new CC problem and to the 
problem of backreaction associated with Hawking radiation,
are qualitatively discussed in Sec.~\ref{DISC}. 

In the paper, we use units in which $\hbar=c=1$, while the
signature of spacetime metric is $(+---)$.

\section{Bohmian semiclassical gravity in first quantization}

\subsection{Bohmian particle trajectories}

Consider the Klein-Gordon equation for a 
massive spin-0 particle in curved spacetime
\begin{equation}\label{KG}
(\nabla^{\mu}\partial_{\mu}+m^2)\psi(x)=0 ,
\end{equation}
where $\nabla^{\mu}$ is the covariant derivative, and the fact that
$\nabla_{\mu}\psi=\partial_{\mu}\psi$ is used.
Eq.~(\ref{KG}) implies the local conservation law
\begin{equation}\label{KGcons}
\nabla^{\mu} \left( \frac{i}{2}
\psi^* \!\stackrel{\leftrightarrow\;}{\partial_{\mu}}\! \psi \right) =0 ,
\end{equation}
which implies that the norm
\begin{equation}\label{norm}
(\psi,\psi) = \int_{\Sigma} d\Sigma^{\mu} \, \frac{i}{2}
\psi^* \!\stackrel{\leftrightarrow\;}{\partial_{\mu}}\! \psi 
\end{equation}
(where $d\Sigma^{\mu}=d^3x \sqrt{|g^{(3)}|} n^{\mu}$ and $n^{\mu}$
is a unit vector normal to  $\Sigma$)
does not depend on the choice of the spacelike hypersurface $\Sigma$.
We consider a solution $\psi$ for which the norm (\ref{norm})
is {\em positive} and equal to 1.

By writing $\psi=Re^{iS}$, where $R$ and $S$ are real functions,
the complex Klein-Gordon equation (\ref{KG}) is equivalent to 
two real equations
\begin{equation}\label{cons}
\nabla^{\mu}(R^2\partial_{\mu}S)=0 ,
\end{equation}
\begin{equation}\label{HJ}
-\frac{(\partial^{\mu}S)(\partial_{\mu}S)}{2m}+\frac{m}{2}+Q=0,
\end{equation}
where
\begin{equation}\label{Q}
Q\equiv \frac{1}{2m}\frac{\nabla^{\mu}\partial_{\mu}R}{R}
\end{equation}
is the quantum potential. Eq.~(\ref{cons}) is the conservation
equation (\ref{KGcons}). Thus, the fact that (\ref{norm})
is unit can be written as 
\begin{equation}
\int d^3x  \sqrt{|g^{(3)}|} R^2\omega =1 ,
\end{equation}
where $\omega(x)=-n^{\mu}(x)\partial_{\mu}S(x)$ is the 
``local frequency". This shows that $R^2\omega$ can be interpreted 
as a probability density of particle positions, provided 
that $R^2\omega$ is non-negative. 
(For the case in which it is locally negative, 
see \cite{nikolfol}.)
Eq.~(\ref{HJ}) can be viewed as a quantum Hamilton-Jacobi equation,
differing from the classical relativistic Hamilton-Jacobi equation
in containing an additional $Q$-term.
Indeed, in physical units with $\hbar\neq 1$, the right-hand side
of (\ref{Q}) attains an additional factor $\hbar^2$, which shows
that $Q\rightarrow 0$ in the classical limit.

In the Bohmian interpretation of relativistic
QM, the particle is a pointlike object having a continuous
trajectory $X^{\mu}(s)$ satisfying the deterministic equation
\cite{durr,nikfpl1,nikfpl3}
\begin{equation}\label{Bohm}
\frac{dX^{\mu}(s)}{ds}=-\frac{1}{m}\partial^{\mu}S ,
\end{equation}
where it is understood that the right-hand side is evaluated
at $x=X$ and
$s$ is an affine parameter along the trajectory.
Using the identity 
\begin{equation}\label{id1}
\frac{d}{ds}=\frac{dX^{\mu}}{ds} \partial_{\mu} ,
\end{equation}
as well as Eqs.~(\ref{Bohm}) and (\ref{HJ}), one finds the 
equation of motion
\begin{equation}\label{beom}
m\frac{D^2X^{\mu}}{Ds^2}= \partial^{\mu}Q ,
\end{equation}
where
\begin{equation}
\frac{D^2X^{\mu}}{Ds^2} \equiv
\frac{d^2X^{\mu}}{ds^2}+\Gamma^{\mu}_{\alpha\beta}
\frac{dX^{\alpha}}{ds}\frac{dX^{\beta}}{ds} .
\end{equation}
The right-hand side of (\ref{beom})
describes the quantum force, i.e., the deviation 
of the particle trajectory from a motion along a geodesic.

\subsection{Energy-momentum tensor}

To construct the conserved energy-momentum tensor associated
to the particle equation of motion (\ref{beom}), we use the 
methods developed in \cite{wein}. The energy-momentum tensor
written in a manifestly covariant form turns out to be
\begin{eqnarray}\label{emt}
T^{\mu\nu}(x) & = & \int ds \frac{\delta^4(x-X(s))}{\sqrt{|g(x)|}} 
\nonumber \\
& & \times \left[ m \frac{dX^{\mu}}{ds} \frac{dX^{\nu}}{ds} -
g^{\mu\nu}(x) Q(x) \right] .
\end{eqnarray}
For a timelike trajectory $X^{\mu}(s)$,
the physical meaning of (\ref{emt}) is more manifest when 
coordinates are chosen such that
$g_{0i}=0$ and $X^0(s)=s/\sqrt{g_{00}}$. 
In this case, (\ref{emt}) can be written as
\begin{eqnarray}\label{emt2}  
T^{\mu\nu}(x) & = & \frac{\delta^3({\bf x}-{\bf X}(s))}{\sqrt{|g^{(3)}(x)|}}
\nonumber \\
& & \times \left[ m \frac{dX^{\mu}}{ds} \frac{dX^{\nu}}{ds} -
g^{\mu\nu}(x) Q(x) \right] ,
\end{eqnarray}
which is nonvanishing only along the particle trajectory ${\bf X}(s)$.
Using (\ref{id1}) (see also \cite{wein}) one finds
\begin{eqnarray}
& \nabla_{\nu} \displaystyle
\int ds \frac{\delta^4(x-X(s))}{\sqrt{|g(x)|}}
m \frac{dX^{\mu}}{ds} \frac{dX^{\nu}}{ds} & \nonumber \\
& = \displaystyle \int ds \frac{\delta^4(x-X(s))}{\sqrt{|g(x)|}}
m \frac{D^2X^{\mu}}{Ds^2} , &
\end{eqnarray}
\begin{eqnarray}
& \nabla_{\nu} \displaystyle
\int ds \frac{\delta^4(x-X(s))}{\sqrt{|g(x)|}}
g^{\mu\nu}(x) Q(x) & \nonumber \\
& = \displaystyle \int ds \frac{\delta^4(x-X(s))}{\sqrt{|g(x)|}}
\partial^{\mu}Q(x) . & 
\end{eqnarray}
Thus, when the equation of motion (\ref{beom}) is satisfied,
then the energy-momentum tensor (\ref{emt}) is conserved:
\begin{equation}
\nabla_{\nu} T^{\mu\nu}(x)=0 .
\end{equation} 
Therefore, it is consistent to introduce a semiclassical 
Einstein equation as
\begin{equation}\label{einst2}
G_{\mu\nu}(x)=8\pi G_{\rm N}T_{\mu\nu}(x).
\end{equation}

Note that the definition of $T^{\mu\nu}$ as above in terms
of pointlike particles is not 
in spirit of the usual formulation of QM. Nevertheless,
assuming that one does not know the actual position of the 
particle, one may obtain an expression more in spirit
of the usual formulation of QM by averaging
over all possible particle positions. Assuming
that $\psi$ is a wave packet localized within a small 3-volume
$\sigma\subset\Sigma$, one makes the replacement
\begin{equation} 
T^{\mu\nu} \rightarrow \langle T^{\mu\nu} \rangle ,
\end{equation}
where $\langle T^{\mu\nu} \rangle$ is the energy-momentum averaged
over the unknown particle positions and attributed to the small region
$\sigma$. The average energy-momentum $\langle T^{\mu\nu} \rangle$ 
is obtained from $T^{\mu\nu}$ in (\ref{emt2}) by making a replacement
\begin{equation}\label{replace}
\frac{\delta^3({\bf x}-{\bf X})}{\sqrt{|g^{(3)}(x)|}} \rightarrow
\frac{1}{v} \int_\sigma d^3x \sqrt{|g^{(3)}(x)|} R^2(x)\omega(x) ,
\end{equation}
where $v\equiv \int_\sigma d^3x \sqrt{|g^{(3)}|}$
and $dX^{\mu}/ds$ is replaced by $-m^{-1}\partial^{\mu}S$,
due to (\ref{Bohm}). Note, however, that the semiclassical 
Einstein equation with such an averaged energy-momentum  
is not physically viable when $\psi$ is not a localized wave packet.
For example, if $\psi$ is a superposition that corresponds 
to {\em two} macroscopically separated lumps, then such a 
semiclassical Einstein equation with an energy-momentum averaged
over both lumps contradicts experiments \cite{page}.
This indicates that the gravitational field responds to the actual 
(not to the average) particle position, so, in general, 
Eq.~(\ref{einst2}) seems more viable as a satisfying 
semiclassical theory of gravity.  

\subsection{Generalization to the many-particle case}

Let us also briefly generalize the results above to the case 
of $n$ particles with mass $m$ described by a wave function
$\psi_n(x_1,\ldots,x_n)$. The wave function satisfies the many-particle 
generalization of (\ref{KG})
\begin{equation}\label{KGn}
\sum_{a=1}^{n} \left( \nabla_a^{\mu}\partial_{a\mu}+m^2
\right) \psi_n(x_1,\ldots,x_n)=0 .
\end{equation}
Thus, all equations above generalize in a trivial way 
by adding an additional label $a$. In particular,
(\ref{Q}) generalizes to
\begin{equation}\label{Qn}
Q_n = \frac{1}{2m} \displaystyle
\frac{ \displaystyle\sum_{a=1}^n \nabla_a^{\mu}\partial_{a\mu}R_n}{R_n} ,
\end{equation}
(\ref{beom}) generalizes to
\begin{equation}\label{beomn}
m\frac{D^2X_a^{\mu}}{Ds^2}= \partial_a^{\mu}Q_n ,
\end{equation}
and (\ref{emt}) generalizes to
\begin{eqnarray}\label{emtn}
T_n^{\mu\nu}(x) & = & \sum_{a=1}^n 
\int ds \frac{\delta^4(x-X_a(s))}{\sqrt{|g(x)|}}
\nonumber \\  
& & \times \left[ m \frac{dX_a^{\mu}}{ds} \frac{dX_a^{\nu}}{ds} -
g^{\mu\nu}(x) Q_n(x) \right] .
\end{eqnarray}
This provides a semiclassical theory of gravity for the case 
in which the number of particles $n$ is fixed.
However, to consider the possibility of particle creation
and destruction, first quantization is not sufficient.
The processes of particle creation and destruction can be described 
by QFT, which we do in the next setion.

\section{Bohmian semiclassical gravity in QFT}
\label{QFT}

\subsection{Particles from QFT}

As an example,
consider a real field $\phi$ in curved spacetime with a 
self-interaction described by the interaction Lagrangian density 
$-(\lambda/4!)\phi^4$.
In the Heisenberg picture, the field operator $\hat{\phi}(x)$ 
satisfies 
\begin{equation}\label{KGf}
\nabla^{\mu}\partial_{\mu}\hat{\phi}(x)+m^2\hat{\phi}(x)
+\frac{\lambda}{3!}\hat{\phi}^3(x) =0 .
\end{equation} 
As outlined in the Introduction and references
cited therein, we assume that 
a preferred foliation of spacetime defines a preferred
notion of particles.
Therefore, an arbitrary QFT state $|\Psi\rangle$
can be written as a superposition of n-particle states as
\begin{equation}\label{Psi}
|\Psi\rangle =\sum_{n=0}^{\infty} c_n |\Psi_n\rangle ,
\end{equation}  
where $|\Psi_n\rangle$ is a normalized $n$-particle state. 
The normalized $n$-particle wave function
is then defined as \cite{schweber,nikfpl1}
\begin{eqnarray}\label{wf}
\psi_n(x_1,\ldots,x_n) & = & \frac{S_{ \{x_a\} }}{\sqrt{n!}} 
\langle 0|\hat{\phi}(x_1)\cdots\hat{\phi}(x_n)|\Psi_n\rangle 
\nonumber \\
&  =  & \frac{S_{ \{x_a\} }}{c_n \sqrt{n!}} 
\langle 0|\hat{\phi}(x_1)\cdots\hat{\phi}(x_n)|\Psi\rangle ,
\end{eqnarray}
where $|0\rangle\equiv |\Psi_0\rangle$ and
$S_{ \{x_a\} }$ denotes the symmetrization over all 
$x_a$, $a=1,\ldots,n$, which is needed because the field 
operators do not commute for nonequal times.
For $\lambda=0$, Eq.~(\ref{KGf}) implies that the wave function
(\ref{wf}) satisfies the $n$-particle Klein-Gordon equation
(\ref{KGn}). 
To see an effect of the self-interaction term in (\ref{KGf})
on the wave functions, we consider an immediate consequence of 
(\ref{KGf}):
\begin{equation}\label{KGf2}
\langle 0| \left[
\nabla^{\mu}\partial_{\mu}\hat{\phi}(x)+m^2\hat{\phi}(x)
+\frac{\lambda}{3!}\hat{\phi}^3(x) 
\right] |\Psi\rangle =0 .
\end{equation}
Eqs.~(\ref{KGf2}) and (\ref{wf}) then imply
\begin{equation}\label{KGf3}
c_1[\nabla^{\mu}\partial_{\mu}+m^2]\psi_1(x)
+\frac{\lambda}{\sqrt{3!}} c_3 \psi_3(x,x,x)=0 .
\end{equation}
Thus the {\em nonlinear} equation (\ref{KGf}) for the field operator
implies a {\em linear} equation for the wave functions, 
such that the nonlinearity transforms into a linear interaction
between wave functions for different numbers of particles.
Eq.~(\ref{KGf3}) also shows under which conditions the particle
described by $\psi_1$ behaves as a free particle
satisfying the free Klein-Gordon equation (\ref{KG}); the interaction 
is non-negligible only when all 4 particles (1 particle described by 
$\psi_1(x)$ and 3 particles described by $\psi_3(x_1,x_2,x_3)$)
are ``close to each other", in the sense that the wave packets
described by
$\psi_1$ and $\psi_3$ have a significant overlap. This is, indeed, 
consistent with the phenomenological picture according to which 
particles need to come close to each other in order to interact
by an interaction such as the $-(\lambda/4!)\phi^4$ theory. 

By writing 
\begin{eqnarray}
& \psi_1(x)=R_1(x)e^{iS_1(x)} , & \nonumber \\
& \psi_3(x,x,x)=R_3(x)e^{iS_3(x)} , &
\end{eqnarray}
and, for simplicity, by assuming that $c_3/c_1$ is real,
the complex equation (\ref{KGf3}) is equivalent to a set of 
two real equations
\begin{equation}\label{HJf}
-\frac{(\partial^{\mu}S_1)(\partial_{\mu}S_1)}{2m}+\frac{m}{2}+Q=0,
\end{equation}
\begin{equation}\label{consf}
\nabla^{\mu}(R_1^2\partial_{\mu}S_1)=J ,
\end{equation}
where
\begin{equation}\label{Qf}
Q\equiv \frac{1}{2m} \left[ \frac{\nabla^{\mu}\partial_{\mu}R_1}{R_1}
+\frac{\lambda}{\sqrt{3!}} \frac{c_3 R_3}{c_1 R_1} \cos(S_1-S_3) \right] ,
\end{equation} 
\begin{equation}\label{J}
J\equiv \frac{\lambda}{\sqrt{3!}} \frac{c_3}{c_1}R_1 R_3 \sin(S_1-S_3) .
\end{equation}
The Bohmian particle trajectory associated with the wave function
$\psi_1(x)$ can be introduced in the same way as in (\ref{Bohm})
with $S\rightarrow S_1$, but now with a modified 
quantum potential (\ref{Qf}). Consequently, the 
associated energy-momentum tensor $T_1^{\mu\nu}$ is given by the 
expression (\ref{emt}), in which $Q$ is given by (\ref{Qf}).
In a similar way, it is straightforward to derive a modified expression
for $T_n^{\mu\nu}$ in (\ref{emtn})
for an $n$-particle wave function (\ref{wf}).
(The expression for $Q_n$ in (\ref{Qn}) attains additional terms
proportional to $\lambda$ similar to that in (\ref{Qf}), but 
we do not write them explicitly as the explicit expression 
for general $n$ is rather cumbersome.) 
In this way one can define $T_n^{\mu\nu}$ for any $n\geq 1$, but 
not for $n=0$. The absence of the
$n=0$ term is a simple consequence of the fact that, 
by definition, the energy-momentum is that of particles
(not of fields), so the no-particle-state (the vacuum) 
does not contribute to the energy-momentum. Perhaps a vacuum 
contribution to the energy-momentum could be introduced by 
hand, but here it would be a rather artificial procedure.
This should be contrasted with the usual field-theoretic 
approach where the fields (not the particles) are regarded
as fundamental objects, so that the vacuum contribution appears
naturally in the field energy-momentum tensor, leading to the 
old CC problem. Here, in our approach with particles regarded 
as more fundamental than fields, the old CC problem simply does not appear.
Turning this argument round,
the fact that the measured cosmological constant is many orders of magnitude
smaller than the one predicted by the field energy-momentum 
indicates that the particles (not the fields) might be the fundamental
objects existing in nature. In this picture, quantum fields are 
merely auxiliar mathematical objects useful for calculation
of certain particle processes, such as particle creation and
destruction. (For a somewhat similar view of QFT, see also \cite{wein2}.)  

Note also that
Eq.~(\ref{consf}) indicates that $R_1^2\omega_1$ is not the probability 
density for the particle described by $\psi_1$ when the overlap
with $\psi_3$ is significant. Nevertheless, the probability density
can be calculated in principle by explicitly calculating the trajectories
for a large sample of initial particle positions, provided that
the initial overlap is negligible, so that the initial 
probability density is given by $R_1^2\omega_1$.
 
\subsection{The effects of particle creation and destruction}

To explicitly take into account the effects of particle creation
and destruction, it is more convenient to work in the
Schr\"odinger picture \cite{nikfpl1,long}. In this picture,
the QFT state is denoted as 
$\Psi[\phi;t)$, which is a functional with respect to $\phi({\bf x})$ 
and a function with respect to $t$.
Eq.~(\ref{Psi}) is now written as
\begin{equation}\label{Psi2}
\Psi[\phi;t) =\sum_{n=0}^{\infty} \tilde{\Psi}_n[\phi;t) ,
\end{equation}
where the tilde above $\tilde{\Psi}_n$ denotes that the norm of it 
may be smaller than unit. In the processes of particle creation and
destruction this norm changes with time.
The field $\phi$ may also be interpreted in a Bohmian deterministic
manner \cite{holrep,holbook}.
By writing $\Psi={\cal R}e^{i{\cal S}}$, one finds an expression
analogous to (\ref{Bohm})
\begin{equation}
\frac{\partial \Phi({\bf x},t)}{\partial t}=
\frac{\delta{\cal S}}{\delta\phi({\bf x})} ,
\end{equation}
where it is understood that the right-hand side is evaluated at 
$\phi=\Phi$. The Bohmian effectivity $e_n$ of the $n$-particle sector of 
(\ref{Psi2}) is \cite{nikfpl1}
\begin{equation}
e_n[\Phi;t)=\frac{ |\tilde{\Psi}_n[\Phi;t)|^2 }
{ \displaystyle\sum_{n'=0}^{\infty} |\tilde{\Psi}_{n'}[\Phi;t)|^2 } .
\end{equation}
The effectivity $e_n$ is a number between 0 and 1 and satisfies
$\sum_{n=0}^{\infty} e_n=1$. 
As shown in \cite{nikfpl1}, when the number of particles is measured,
then $e_n$ becomes $e_n=1$ for one $n$ and $e_{n'}=0$ for 
all other $n'$.
This corresponds to an effective collapse of (\ref{Psi2}) to 
one of $\Psi_n$'s, which is induced by the quantum measurement.
The probability for such an effective collapse is exactly
equal to the corresponding probability predicted by the 
standard probabilistic rules of QFT \cite{nikfpl1}. 
However, when the number of particles is not measured, i.e., when 
more than one $e_n$ is different from 0, then all $T_n^{\mu\nu}$
for which $e_n\neq 0$ 
contribute to the total energy-momentum. Thus, the total
energy-momentum is
\begin{equation}\label{ee1}
T^{\mu\nu}=\sum_{n=1}^{\infty} e_n T_n^{\mu\nu} +U^{\mu\nu}. 
\end{equation}
The additional term $U^{\mu\nu}$ is a compensating term 
that provides the conservation of $T^{\mu\nu}$ even when
the effectivities 
$e_n$ change with time. Since $\nabla_{\nu}T_n^{\mu\nu}=0$ by 
construction, the requirement
\begin{equation}
\nabla_{\nu}T^{\mu\nu}=0
\end{equation}
leads to the equation
\begin{equation}\label{ee4}
\nabla_{\nu}U^{\mu\nu}=j^{\mu} ,
\end{equation}
where
\begin{equation}\label{ee4j}
j^{\mu} \equiv -\sum_{n=1}^{\infty} (\partial_{\nu}e_n) T_n^{\mu\nu} .
\end{equation}
We see that $j^{\mu}$ can be viewed as a collection of pointlike 
sources nonvanishing only along the particle trajectories.
However, in (\ref{ee1}) we do not want $U^{\mu\nu}$ 
to be nonvanishing only along the particle trajectories, because 
then $U^{\mu\nu}$ would simply cancel the pointlike 
energy-momentum of new created particles described by the first term in
(\ref{ee1}), so that the new created particles would not influence
the gravitational field. 
Instead, we want equation (\ref{ee4}) to 
describe a {\em continuous} field $U^{\mu\nu}(x)$
produced by the pointlike sources $j^{\mu}$.
This makes $U^{\mu\nu}$ in (\ref{ee4}) similar to the electromagnetic field
described by the Maxwell equations,
but with an important difference consisting in the fact that
$U^{\mu\nu}$ is a symmetric tensor, whereas the electromagnetic field 
is an antisymmetric tensor.
Therefore, we assume
\begin{equation}\label{ansatz}
U^{\mu\nu}=\nabla^{\mu}V^{\nu}+\nabla^{\nu}V^{\mu} ,
\end{equation}
where $V^{\mu}(x)$ is a vector field analogous 
to the electromagnetic potential. Now (\ref{ee4}) becomes
\begin{equation}\label{ee4.1}
\nabla_{\nu} \nabla^{\mu} V^{\nu} + \nabla_{\nu} \nabla^{\nu} V^{\mu}=
j^{\mu} ,
\end{equation}
which describes the propagation of the field $V^{\mu}$, the source of which
is a collection of pointlike sources described by $j^{\mu}$.  
Eq.~(\ref{ee4.1}) represents a set of 4 equations for 4 unknowns
$V^{\mu}$, which further justifies the ansatz (\ref{ansatz}).

In some cases, the solution of (\ref{ee4.1}) can be found explicitly.
For example, assume (i) that spacetime can be approximated by a
flat spacetime and (ii) that $\partial_{\nu}e_n$ 
changes slowly, so that one can use the approximation
$\partial_{\mu}\partial_{\nu}e_n \simeq 0$. In this case, 
(\ref{ee4.1}) can be written as
\begin{equation}\label{ee7}
\partial^{\mu} \partial_{\nu} V^{\nu} + \partial_{\nu} \partial^{\nu} V^{\mu}=
j^{\mu} ,
\end{equation}
while $j^{\mu}$ is approximately conserved:
\begin{equation}\label{ee8}
\partial_{\mu}j^{\mu}=
-\sum_{n=1}^{\infty} (\partial_{\mu}\partial_{\nu}e_n) T_n^{\mu\nu}
\simeq 0.
\end{equation}
Introducing the well-known retarded Green function $G(x-x')$
satisfying
\begin{equation}
\partial_{\nu} \partial^{\nu} G(x-x')=\delta^4(x-x') ,
\end{equation}
the explicit solution of (\ref{ee7}) is
\begin{equation}\label{ee10}
V^{\nu}(x)=\int d^4x'\, G(x-x') j^{\nu}(x') .
\end{equation}
Indeed, (\ref{ee8}) implies that (\ref{ee10}) satisfies the 
Lorentz condition
\begin{equation}\label{ee8.1}
\partial_{\nu}V^{\nu}(x)=
\int d^4x'\, G(x-x') \partial'_{\nu}j^{\nu}(x') \simeq 0 ,
\end{equation}
so (\ref{ee7}) reduces to 
$\partial_{\nu} \partial^{\nu} V^{\mu}=j^{\mu}$, which, indeed, 
is satisfied by (\ref{ee10}). 

Now the final semiclassical Einstein equation reads
\begin{equation}\label{semiclfin}
G_{\mu\nu}(x)=8\pi G_{\rm N} T_{\mu\nu}(x) ,
\end{equation}
where the quantum matter energy-momentum tensor $T_{\mu\nu}(x)$ 
is given by (\ref{ee1}). Of course, we have explicitly analyzed
only the contributions from massive spinless uncharged particles
corresponding to the hermitian field $\hat{\phi}$, but the contributions
from other types of particles can be introduced in a similar way.
Some additional physical features of the resulting semiclassical theory
are qualitatively discussed in the next section.   

\section{Discussion and conclusion}
\label{DISC}

As we have seen, by regarding particles as more fundamental objects 
than fields, the usual field energy-momentum tensor no longer 
represents the physical energy-momentum, which automatically 
solves (or at least avoids) the old CC problem, simply because
only particles contribute to the physical energy-momentum.
However, it is important to emphasize that, 
by discarding the field ground-state energy,
we do {\em not} discard the {\em particle} ground-state energy.
The QFT ground state containing no particles is physically 
very different from the particle ground state. The best known
example of the latter is a single particle in a one-dimensional 
harmonic-oscillator potential $V(x)=m\omega^2 x^2/2$, 
where the ground state having the nonrelativistic energy $\omega/2$
is still a {\em one-particle} (not zero-particle) state. Indeed, such a 
particle ground-state energy is included in the particle 
energy-momentum (\ref{emt}). In fact, the second term 
in (\ref{emt}) proportional to $g^{\mu\nu}Q$ is exactly 
of the form of a cosmological term. Moreover, in a nonrelativistic
limit one may expect that 
$\partial^{\mu}\partial_{\mu} R \sim \pm m^2 R$, so (\ref{Q}) implies
\begin{equation}
|Q|\sim m .
\end{equation}
This means that particles with a mass $m$ may contribute to the 
cosmological constant by a contribution of the order of $mn_v$,
where $n_v$ is the number of particles per unit volume.
It is tempting to speculate that this could have something to do
with the coincidence problem, i.e., with the {\em new} CC problem.
Note, however, that a plane wave $e^{-ik\cdot x}$ has a 
constant $R$, so $(\ref{Q})$ vanishes for a plane wave.
Nevertheless, it is conceivable that the so-called 
dark energy might consist of particles described by 
a nontrivial wave function that leads to a nontrivial quantum
potential $Q$, so that (i) the energy-momentum of these particles is 
dominated by a cosmological term $\propto g^{\mu\nu}Q$ and 
(ii) the quantum force described by (\ref{beom}) prevents 
these particles from forming structures.
Such a wave function should be a wave packet with a width larger
than typical scales associated to cosmological structures.
(The needed large width might be a natural consequence of inflation.)
However, a more serious investigation
of such a possibility would require a further theoretical input,
which would go beyond the scope of the present paper.

Concerning the issue of the new CC problem, we recall that
a term proportional to $\lambda$ also survives in (\ref{Qf}).
This demonstrates that a nontrivial field potential may also 
influence the cosmological constant. In particular, 
it means that 
the quintessence models of dark energy may also play a role
for the new CC problem, provided that they are reinterpreted 
in terms of particle wave functions, analogously to that in (\ref{KGf3}).
A similar remark applies also to scalar-field potentials
supposed to drive the early cosmological inflation.

Another new physical ingredient that we want to discuss is the 
physical meaning of $U^{\mu\nu}$ in (\ref{ee1}). 
Unlike the first term in (\ref{ee1}), $U^{\mu\nu}$ represents 
a continuously distributed contribution to the total energy-momentum.
Thus, it is a nonparticle contribution to the energy-momentum, 
but the particles are the source for it. More precisely,
from (\ref{ee4}) and (\ref{ee4j}) we see that
$U^{\mu\nu}$ is created only when the effectivities $e_n$ 
change with time. Physically, this means that a particle 
that gets destroyed compensates it by emitting positive
$U$-energy, while a particle 
that gets created compensates it by emitting negative
$U$-energy.
In fact, in most physical processes
with particle creation and destruction (usually described 
by the S-matrix formalism in elementary-particle physics) 
the energy-momentum of the initial particles is exactly equal 
to the energy-momentum of the final particles. This means that
$U^{\mu\nu}$ averaged over a large volume vanishes
in the initial as well as in the final state
of such a process. The creation of $U^{\mu\nu}$ as described by 
(\ref{ee4}) is only      
a transient phenomenon, not directly observable in 
typical particle collision and decay processes.
On the other hand, when particles are created from an unstable 
{\em vacuum}, then the conservation of $T^{\mu\nu}$ implies 
that average $U^{\mu\nu}$ must be nonzero even in the final state.
In particular, this provides a backreaction mechanism 
for the process of Hawking radiation, in which particles
are created from the vacuum in a background of a classical
black-hole \cite{bd}. Thus, Eq.~(\ref{semiclfin}) may
be applied to a new analysis of the process of Hawking radiation
with backreaction, but a detailed analysis of such a
process is beyond the scope of the present paper.  
It is also fair to note that the ansatz (\ref{ansatz}) 
is not necessarily the only possibility.

To conclude, the formulation of semiclassical gravity in terms
of Bohmian particle trajectories has several advantages over 
the usual formulation. First, regarding particles (rather than
fields) as the fundamental physical objects automatically 
avoids the old CC problem. Second, the use of the Bohmian
formulation of quantum theory avoids the discontinuous collapse 
problem. Besides, this formulation suggests new approaches
to the solution of the new CC problem and of the backreaction 
problem associated to particle creation by classical gravitational fields.
Thus, we believe that our new approach to semiclassical
gravity is worthwhile of further investigation. 

\section*{Acknowledgments}
The author is grateful to Z. Zakir for valuable remarks.
This work was supported by the Ministry of Science and Technology of the
Republic of Croatia.

\end{document}